\documentclass[12pt]{article}
\usepackage{epsfig}
\usepackage{latexsym}
\usepackage{graphicx}
\usepackage{amsmath}
\usepackage{amsfonts}
\usepackage{amssymb}

\newtheorem{theorem}{Theorem}

\newtheorem{lemma}[theorem]{Lemma}
\newtheorem{corollary}[theorem]{Corollary}
\newtheorem{conjecture}[theorem]{Conjecture}

\newtheorem{observation}[theorem]{Observation}

\newcommand{\qed}{\hfill\rule{0.5em}{0.809em}}

\newenvironment{proof}{
\par
\noindent {\bf Proof.}\rm}{\mbox{}\hfill\rule{0.5em}{0.809em}
\par\mbox{}\par}

\begin{document}
\title{Coloring the square of the\\
 Cartesian product of two cycles}
\author{\'Eric Sopena\thanks{E-mail: sopena@labri.fr.}$\ $ and
        Jiaojiao Wu\thanks{E-mail: wujj0007@yahoo.com.tw.} \\
        \\
  {\small Universit\'e de Bordeaux, LaBRI UMR 5800,}\\
  {\small 351, cours de la Lib\'eration,}\\
  {\small F-33405 Talence  Cedex, France}
}
\date{\today}

\maketitle

\begin{abstract}
The square $G^2$ of a graph $G$ is defined on the vertex set of $G$ in such a way
that distinct vertices with distance at most 2 in $G$ are joined by an edge.
We study the chromatic number of the square of the Cartesian product $C_m\Box C_n$ of
two cycles and show that the value of this parameter is at most 7 except
when $(m,n)$ is $(3,3)$, in which case the value is 9, and when $(m,n)$ is $(4,4)$ or $(3,5)$,
in which case the value is 8.
Moreover, we conjecture that whenever $G=C_m\Box C_n$, the chromatic number of
$G^2$ equals $\lceil mn/\alpha(G^2) \rceil$,
where $\alpha(G^2)$ denotes the maximum size of an independent set in $G^2$.
\end{abstract}


\section{Introduction}

A {\em proper $k$-coloring} of a graph $G$ with vertex set $V(G)$
and edge set $E(G)$ is a mapping $c$ from $V(G)$
to the set $\{1,2,\dots,k\}$ such that $c(u)\neq c(v)$ whenever $uv$ is an edge in $E(G)$.
The {\em chromatic number} $\chi(G)$ of $G$ is the smallest $k$ for which $G$ admits a proper $k$-coloring.

Let $G$ and $H$ be graphs. The {\em Cartesian product} $G \Box H$ of $G$ and $H$
is the graph with vertex set $V(G) \times V(H)$ where two vertices
$(u_1, v_1)$ and $(u_2,v_2)$ are adjacent if and only if either
$u_1=u_2$ and $v_1v_2 \in E(H)$ or $v_1=v_2$ and $u_1u_2 \in E(G)$.
Let $P_n$ and $C_n$ denote respectively the path and the cycle on $n$ vertices.
We will denote by $G_{m,n}$ the {\em grid} $P_m\Box P_n$ with $m$ rows and $n$ columns
and by $T_{m,n}$ the {\em toroidal grid} $C_m\Box C_n$ with $m$ rows and $n$ columns.

The {\em square} $G^2$ of a graph $G$ is given by
$V(G^2)=V(G)$ and $uv\in E(G^2)$ if and only if $uv\in E(G)$ or $u$ and $v$
have a common neighbor.
In other words, any two vertices within distance at most~2 in $G$ are linked by an edge in $G^2$.
Let $\Delta(G)$ denote the maximum degree of $G$.
The problem of determining the chromatic number of the square of particular graphs has
attracted very much attention, with a particular focus on the square of planar graphs
(see e.g.~\cite{DKNS,HHMR,HM,LW,MS}), following Wegner~\cite{W}, who conjectured
that every planar graph $G$ with maximum degree at least 8 satisfies
$\chi(G^2)\le\lfloor\frac{3}{2}\Delta(G)\rfloor+1$.
Havet {\em et al.} proved in~\cite{HHMR} that the square of any such planar graph
admits a proper coloring using $(\frac{3}{2}+o(1))\Delta(G)$ colors.

In~\cite{CY}, Chiang and Yan studied the chromatic number of the square of Cartesian
products of paths and cycles and proved the following:

\begin{theorem}[Chiang and Yan~\cite{CY}]
If $G=C_m \Box P_n$ with $m\ge3$ and $n\ge 2$,  then
$$\chi(G^2)=\left \{
\begin{array}{ll}
 4 & \quad \mbox{if $n=2$ and $m \equiv 0 \pmod 4$,}\\
 6 & \quad \mbox{if $n=2$ and $m\in\{3,6$\},}\\
 6 & \quad \mbox{if $n \geq 3$ and $m \not\equiv 0 \pmod 5$,}\\
 5 & \quad \mbox{otherwise.}\\
\end{array} \right .
$$
\label{CY-Theorem}
\end{theorem}
 
Since $C_m \Box P_n$ is a subgraph of $C_m \Box C_n$, Theorem~\ref{CY-Theorem} provides
lower bounds for the chromatic number of the square of toroidal grids.

A proper coloring of the square $G^2$ of a graph $G$ is often called
a {\em distance-2 coloring} of $G$.
In~\cite{PW}, P\'or and Wood studied the notion of ${\cal F}$-free coloring.
Let $\cal F$ be a family of connected bipartite graphs, each with at least three
vertices. An {\em ${\cal F}$-free coloring} of a graph $G$ is then a proper vertex coloring of $G$
with no bichromatic subgraph in $\cal F$.
This notion
generalizes several types of colorings and, in particular, distance-2 coloring
when ${\cal F}=\{P_3\}$.
They obtained an upper bound on the ${\cal F}$-free chromatic number
of cartesian products of general graphs. Moreover, in case of distance-2 coloring, they proved
that the chromatic number of the square of any graph given as the Cartesian product
of $d$ cycles is at most $6d+O(\log d)$.

An {\em $L(p,q)$-labeling} of a graph $G$ is an assignment $\phi$ of nonnegative integers to
the vertices of $G$ so that $|\phi(u)-\phi(v)|\ge p$ whenever $u$ and $v$
are adjacent and $|\phi(u)-\phi(v)|\ge q$
whenever $u$ and $v$ are at distance 2 in $G$.
The {\em $\lambda^p_q$-number} of $G$ is defined as the smallest $k$ such that $G$
admits an $L(p,q)$-labeling on the set $\{0,1,\dots,k\}$ (note that such a labeling
uses $k+1$ labels).
It follows from the definition that any $L(1,0)$-labeling of $G$ is an ordinary proper coloring of $G$ and that
any $L(1,1)$-labeling of $G$ is a proper coloring of the square of $G$.
Therefore, $\chi(G)=\lambda^1_0(G)+1$ and $\chi(G^2)=\lambda^1_1(G)+1$ for every graph $G$.

The notion of $L(p,q)$-labeling
was introduced by Griggs and Yeh~\cite{GY} to model the
{\em Channel Assignment Problem}. They conjectured that $\lambda^2_1(G)\le\Delta(G)^2$
for every graph $G$.
This motivated many authors to study $L(2,1)$-labeling of some particular classes of graphs,
and the case of Cartesian products of graphs was investigated in~\cite{CY,GMS,J,JKV,JNSSS,KY,ST,WGM}.

In particular, Schwartz and Troxell~\cite{ST} considered $L(2,1)$-labelings of products of
cycles and proved the following:

\begin{theorem}[Schwartz and Troxell~\cite{ST}]
If $T_{m,n}=C_m \Box C_n$ with $3\le m\le n$, then

$$\begin{array}{rcl}
\lambda^2_1(T_{m,n}) & = &
\left\{ \begin{array}{ll}
  6 & \quad \mbox{if }\  m,n\equiv 0 \pmod 7,\\
  8 & \quad \mbox{if }\  (m,n)\in A, \\
  7 & \quad \mbox{otherwise.}
\end{array}
\right .
\end{array}$$
where $A=\big\{(3,i):\ i\in\{4,10\}\ \mbox{or}\ i=2j+1\ \mbox{with}\
j\in\mathbb{N}\big\} \cup
\big\{(5,i):\ i\in\{5,6,9,10,13,17\}\big\} \cup
\big\{(6,7),(6,11),(7,9),(9,10)\big\}$.
\label{ST-Theorem}
\end{theorem}

Since every $L(2,1)$-labeling is an $L(1,1)$-labeling,
$\lambda^2_1(G)+1\ge\lambda^1_1(G)+1=\chi(G^2)$ for every graph $G$.
Therefore, Theorem~\ref{ST-Theorem} provides upper bounds on the
chromatic number of the square of toroidal grids (the
upper bounds corresponding to the three cases of Theorem~\ref{ST-Theorem}
are 7, 9, and 8, respectively).

Our main result will improve the bounds provided by Theorems~\ref{CY-Theorem}
and~\ref{ST-Theorem} and by the general result of P\'or and Wood~\cite{PW}:

\begin{theorem}
If $T_{m,n}=C_m \Box C_n$ with $3\le m\le n$, then
$\chi(T_{m,n}^2) \leq 7$ except $\chi(T_{3,3}^2)=9$ and
$\chi(T_{3,5}^2)=\chi(T_{4,4}^2)=8$.
\label{SW-Theorem}
\end{theorem}

\section{Coloring the squares of toroidal grids}
In this section, we shall prove Theorem~\ref{SW-Theorem}
and give more precise bounds for Cartesian products
of some particular cycles.

We shall construct explicit colorings using combinations of {\em patterns}
given in matrix form. Each pattern can be thought of as a proper coloring of the square
of the toroidal grid of the same size. For instance, the pattern $E$ depicted in
Figure~\ref{Figure-CDE} provides in an obvious way a proper $7$-coloring of the square
of $T_{3,7}$. Moreover, by repeating this pattern, we can easily obtain
a proper $7$-coloring of the square of toroidal grids of the form $T_{3m,7q}$.

Let $G$ be a graph and $c$ be a proper coloring of $G$. Since every color class
under $c$ is an independent set, we have the following standard observation:

\begin{observation}\label{lowerbound}
$\chi(G) \geq \Big\lceil\frac{|V(G)|}{\alpha(G)}\Big\rceil$ where $\alpha(G)$ denotes the
maximum size of an independent set in $G$.
\label{Obs-lowerbound}
\end{observation}

We shall extensively use a result of Sylvester.
Given two integers $r$ and $s$, let $S(r,s)$ denote the set of all nonnegative integer combinations of $r$ and $s$:
 $$S(r, s)=\{\alpha r + \beta s : \alpha, \beta \mbox{ nonnegative integers}\}.$$

\begin{lemma}[Sylvester]
If $r$ and $s$ are relatively prime integers greater than~1, then $t \in S(r, s)$ for
all $\ t \geq (r-1)(s-1)$, and $(r-1)(s-1)-1 \notin S(r, s)$.
\label{Lemma-rs}
\end{lemma}

We then have:

\begin{theorem}
If $T_{m,n}=C_m \Box C_n$ with  $m \in S(4, 7)$ and $n \in S(3, 7)$, then $\chi(T_{m,n}^2) \leq 7$.
\label{Theorem-m4n3}
\end{theorem}

\begin{proof}
Let $m \in S(4, 7)$ and $n \in S(3, 7)$. We use the following $7 \times 7$ pattern $A$
to prove the claim.

$$
A=
  \begin{array}{|ccccccc|}
    \hline
    1& 6& 4& 2& 7& 5& 3\\
    2& 7& 5& 3& 1& 6& 4\\
    3& 1& 6& 4& 2& 7& 5\\
    4& 2& 7& 5& 3& 1& 6\\
    5& 3& 1& 6& 4& 2& 7\\
    6& 4& 2& 7& 5& 3& 1\\
    7& 5& 3& 1& 6& 4& 2\\
    \hline
\end{array}
$$

\mbox{}\par

It is easy to check that this pattern properly colors $T_{7,7}^2$.
For any pattern $X$, let $X_i$, $X'_j$ be the subpatterns of $X$
such that $X_i$ is obtained by taking the $i$ first rows of $X$ and $X'_j$ is
obtained by taking the $j$ first columns of $X$.
It is again easy to check that the patterns $A_4$ and $A'_3$ provide
proper colorings of $T_{4,7}^2$ and $T_{7,3}^2$, respectively.
Therefore, using combinations of $A$ and $A_4$, we can
get a $m \times 7$ pattern  $Y$. Moreover, using combinations of $Y$ and $Y'_3$, we can get
a $m \times n$ pattern that provides a proper $7$-coloring of $T_{m,n}^2$,
except when $(m,n)=(7a+4b,7c+3d)$ with $a,c\ge 0$ and $b,d>0$.
In that case, it is enough to replace the color 4 in the upper-right corner
of the rightmost copy of $Y'_3$ by 3 (see example below).
\end{proof}

For example, the following pattern $B$ provides a proper $7$-coloring of $T_{11,13}^2$,
obtained from $A$ by using the combinations
$11=7+4$ and $13=7+2\times3$ and replacing the color 4 by 3 in the upper-right corner.

$$B=
  \begin{array}{|ccccccc|ccc|ccc|}
    \hline
    1& 6& 4& 2& 7& 5& 3& 1& 6& 4& 1& 6& {\bf\underline{3}}   \\
    2& 7& 5& 3& 1& 6& 4& 2& 7& 5& 2& 7& 5 \\
    3& 1& 6& 4& 2& 7& 5& 3& 1& 6& 3& 1& 6 \\
    4& 2& 7& 5& 3& 1& 6& 4& 2& 7& 4& 2& 7\\
    5& 3& 1& 6& 4& 2& 7& 5& 3& 1& 5& 3& 1\\
    6& 4& 2& 7& 5& 3& 1& 6& 4& 2& 6& 4& 2\\
    7& 5& 3& 1& 6& 4& 2& 7& 5& 3& 7& 5& 3\\
    \hline
    1& 6& 4& 2& 7& 5& 3& 1& 6& 4& 1& 6& 4   \\
    2& 7& 5& 3& 1& 6& 4& 2& 7& 5& 2& 7& 5 \\
    3& 1& 6& 4& 2& 7& 5& 3& 1& 6& 3& 1& 6 \\
    4& 2& 7& 5& 3& 1& 6& 4& 2& 7& 4& 2& 7\\
    \hline
     \end{array}
$$

\mbox{}\par

By Lemma~\ref{Lemma-rs} we then get:

\begin{corollary}
If $T_{m,n}=C_m \Box C_n$ with  $m \geq 12$ and $n \geq 18$,
then $\chi(T_{m,n}^2) \leq 7$.
\label{m12n18}
\end{corollary}

We now consider $T^2_{3,n}$. 

\begin{theorem}
If $T_{3,n}=C_3 \Box C_n$ with $n\ge 3$,
then $$\chi (T_{3,n}^2)=\left \{
\begin{array}{ll}
 6 & \quad \mbox{if $n$ is even,}\\
 7 & \quad \mbox{if $n$ is odd and $n\ge 7$,}\\
 8 & \quad \mbox{if $n=5$,}\\
 9 & \quad \mbox{if $n=3$.}\\
\end{array} \right .$$
\label{Theorem-3n}
\end{theorem}

\begin{figure}

$$\begin{array}{ccc}
  C=
  \begin{array}{|cccc|}
     \hline
     1& 4& 2 & 5\\
     2& 5& 3 &6\\
     3& 6& 1 & 4\\
     \hline
  \end{array}

&
  \ D=
  \begin{array}{|cccccc|}
     \hline
     1& 4& 2& 5& 3& 6 \\
     2& 5& 3& 6& 1& 4\\
     3& 6& 1& 4& 2& 5\\
     \hline
  \end{array}
&
  \ E=
  \begin{array}{|ccccccc|}
     \hline
     1& 4& 2& 3& 1& 2& 5 \\
     2& 5& 1& 4& 7& 3& 6\\
     3& 6& 7& 5& 6& 4& 7\\
     \hline
  \end{array}
\\
\end{array}$$
\caption{Patterns for Theorem~\ref{Theorem-3n}\label{Figure-CDE}}
\end{figure}

\begin{proof}
Let $C$, $D$, and $E$ be the patterns given in Figure~\ref{Figure-CDE}.
These patterns clearly provide proper colorings of $T_{3,4}^2$,
$T_{3,6}^2$, and $T_{3,7}^2$, respectively.
For the upper bounds, we use the combinations of patterns $C$ and $D$ to obtain the even cases
and use the combinations of patterns $C$, $D$, and $E$ to obtain the odd cases.
The remaining cases are $n\in\{3, 5, 9\}$, and the following patterns provide
the required proper colorings of $T_{3,3}$, $T_{3,5}$, and $T_{3,9}$, respectively.

$$
  \begin{array}{|ccc|}
     \hline
     1& 4& 7 \\
     2& 5& 8 \\
     3& 6& 9 \\
     \hline
  \end{array}
\ \ \
  \begin{array}{|ccccc|}
     \hline
     1& 4& 2& 3& 6 \\
     2& 5& 1& 4& 7 \\
     3& 6& 7& 5& 8 \\
     \hline
  \end{array}
\ \ \
  \begin{array}{|ccccccccc|}
     \hline
     1& 4& 2& 3& 1& 4& 5& 3& 6  \\
     2& 5& 1& 4& 2& 3& 6& 4& 7 \\
     3& 6& 7& 5& 6& 7& 1& 2& 5 \\
     \hline
     \end{array}
$$

\mbox{}\par

For the lower bounds, notice that the intersection of any independent set $I$ in
$T^2_{3,n}$ with any two consecutive columns contains at most one vertex.
Therefore, $\alpha(T_{3,n}^2) \leq \lfloor n/2\rfloor$.
By Observation~\ref{Obs-lowerbound}, $\chi(T_{3,n}^2)>6$ when $n$ is odd; also,
$\chi(T_{3,n}^2)>7$ when $n=5$ and $\chi(T_{3,n}^2)\geq 9$ when $n=3$.
\end{proof}

As in the proof of Theorem~\ref{Theorem-m4n3}, we can obtain proper colorings
of $T_{3k,n}^2$, for $k\ge 1$, by using combinations of the patterns given in
Theorem~\ref{Theorem-3n}. We thus get the following:

\begin{corollary}\label{3k}
If $T_{3k,n}=C_{3k} \Box C_n$ with $k\ge 1$ and $n\ge 3$, then
$$\chi (T_{3k,n}^2)\leq \left \{
\begin{array}{ll}
 6 & \quad \mbox{if $n$ is even,}\\
 7 & \quad \mbox{if $n$ is odd and $n\ge 7$,}\\
 8 & \quad \mbox{if $n=5$,}\\
 9 & \quad \mbox{if $n=3$.}\\
\end{array} \right .$$
\end{corollary}

We now consider $T^2_{4,n}$.

\begin{theorem}
If $T_{4,n}=C_4 \Box C_n$ with $n\ge 3$, then
$$\chi (T_{4,n}^2)=\left \{
\begin{array}{ll}
 6 & \quad \mbox{if $n\equiv 0 \pmod 3$,}\\
 8 & \quad \mbox{if $n=4$,}\\
 7 & \quad \mbox{otherwise.}\\
\end{array} \right .$$
\label{Theorem-4n}
\end{theorem}

\begin{figure}
$$
\begin{array}{ccc}
F=
  \begin{array}{|ccc|}
     \hline
     1& 3& 5 \\
     2& 4& 6 \\
     3& 5& 1 \\
     4& 6& 2 \\
     \hline
\end{array}
& \ \ \ &
G=
  \begin{array}{|ccccc|}
     \hline
     1& 3& 2& 4& 6 \\
     2& 4& 6& 3& 5 \\
     3& 5& 7& 2& 1 \\
     4& 6& 1& 5& 7 \\
     \hline
   \end{array}
\\
\\
H_1=
  \begin{array}{|cccc|}
     \hline
     1& 2& 3& 4 \\
     3& 4& 5& 6 \\
     5& 6& 7& 8 \\
     7& 8& 1& 2 \\
     \hline
  \end{array}
& \ \ \ &
H_2=
  \begin{array}{|ccccccc|}
     \hline
     1& 3& 2& 6& 4& 7& 5 \\
     2& 4& 5& 7& 1& 3& 6 \\
     3& 1& 6& 2& 5& 4& 7 \\
     4& 5& 7& 1& 3& 6& 2 \\
     \hline
  \end{array}
\\
\end{array}
$$
\caption{Patterns for Theorem~\ref{Theorem-4n}\label{Figure-FGH}}
\end{figure}

\begin{proof}
For $m=3k$, this follows from Corollary~\ref{3k}.
Let $F$ and $G$ be the patterns given in
Figure~\ref{Figure-FGH}. These patterns clearly provide proper colorings
of $T_{4,3}^2$ and $T_{4,5}^2$, respectively.
Thanks to Lemma~\ref{Lemma-rs}, by using combinations of $F$ and $G$, we can get
a proper $7$-coloring of $T_{4,n}^2$ except when $n\in\{4,7\}$.
By using patterns $H_1$ and $H_2$ given in Figure~\ref{Figure-FGH},
we obtain proper colorings of $T_{4,4}^2$ and $T_{4,7}^2$, respectively.

An independent set in
$T^2_{4,n}$ has at most two vertices in any three consecutive columns.
Thus, $\alpha(T_{4,n}^2) \leq \lfloor\frac{2n}{3}\rfloor$. By Observation~\ref{Obs-lowerbound},
$\chi(T_{4,n}^2)>6$ when $n$ is not a multiple of $3$ and $\chi(T_{4,n}^2)\geq 8$ when $n=4$.
\end{proof}

Using combinations of the patterns from Theorem~\ref{Theorem-4n}, we get the following:

\begin{corollary}\label{4k}
If $T_{4k,n}=C_{4k} \Box C_n$ with $k\ge 1$ and $n\ge 3$, then
$$\chi (T_{4k,n}^2)\leq \left \{
\begin{array}{ll}
 6 & \quad \mbox{if $n\equiv 0 \pmod 3$,}\\
 8 & \quad \mbox{if $n=4$,}\\
 7 & \quad \mbox{otherwise.}\\
\end{array} \right .$$
\end{corollary}

We now consider $T^2_{5,n}$.

\begin{theorem}
If $T_{5,n}=C_5 \Box C_n$ with $n \geq 5$, then
$$\chi (T_{5,n}^2)=\left \{
\begin{array}{ll}
 5 & \quad \mbox{if $n\equiv 0 \pmod 5$,}\\
 7 & \quad \mbox{if $n=7$,}\\
 6 & \quad \mbox{otherwise.}\\
\end{array} \right .$$
\label{Theorem-5n}
\end{theorem}

\begin{figure}
$$
\begin{array}{cc}
I=
  \begin{array}{|ccccc|}
    \hline
    1& 2& 3& 4& 5\\
    3& 4& 5& 1& 2\\
    5& 1& 2& 3& 4\\
    2& 3& 4& 5& 1\\
    4& 5& 1& 2& 3 \\
    \hline
\end{array}
&
J=
  \begin{array}{|cccccc|}
     \hline
     6& 1& 2& 3& 4& 5 \\
     3& 4& 5& 6& 1& 2 \\
     5& 6& 1& 2& 3& 4 \\
     2& 3& 4& 5& 6& 1 \\
     4& 5& 6& 1& 2& 3 \\
     \hline
   \end{array}
\\
\end{array}
$$
\caption{Patterns for Theorem~\ref{Theorem-5n}\label{Figure-IJ}}
\end{figure}

\begin{proof}
Let $I$ and $J$ be the patterns given in Figure~\ref{Figure-IJ}; they provide
proper colorings of $T_{5,5}^2$ and $T_{5,6}^2$, respectively.
We use combinations of $I$ and $J$
to get a proper $5$-coloring (resp. $6$-coloring) of $T_{5,n}^2$ when $n\equiv 0 \pmod 5$
(resp. when $n \in S(5,6)$).

The remaining cases are $n\in\{7, 8, 9, 13, 14, 16, 19\}$.
The corresponding patterns are given below,
except for $n=16$, in which case the corresponding pattern is obtained by combining
two $5 \times 8$ patterns.

$$
\begin{array}{cc}
\ \
  \begin{array}{|ccccccc|}
    \hline
    1& 3& 2& 1& 7& 5& 4\\
    2& 4& 5& 3& 6& 1& 7\\
    3& 1& 6& 7& 5& 4& 6\\
    4& 2& 3& 4& 2& 7& 1\\
    5& 6& 7& 5& 3& 6& 2\\
    \hline
\end{array}
\ \
&
\ \
  \begin{array}{|cccccccc|}
    \hline
    1& 3& 2& 6& 3& 1& 5& 4\\
    2& 4& 5& 1& 4& 2& 3& 6\\
    3& 1& 6& 2& 5& 6& 4& 5\\
    4& 2& 3& 4& 1& 3& 2& 1\\
    5& 6& 1& 5& 2& 4& 6& 3\\
    \hline
  \end{array}
\ \ \\
n=7 & n=8
\end{array}
$$

$$
\begin{array}{cc}
\
  \begin{array}{|ccccccccc|}
    \hline
    1& 3& 5& 1& 4& 6& 5& 2& 4\\
    2& 4& 6& 3& 2& 1& 4& 3& 5\\
    3& 1& 2& 4& 6& 5& 2& 1& 6\\
    4& 6& 3& 5& 1& 4& 3& 5& 2\\
    5& 2& 4& 6& 3& 2& 1& 6& 3\\
    \hline
  \end{array}
  \
 &
\
  \begin{array}{|ccccccccccccc|}
    \hline
    1& 3& 2& 6& 3& 1& 5& 2& 4& 3&6&5&4\\
    2& 4& 5& 1& 4& 2& 6& 3& 1&5&2&1&6\\
    3& 1& 6& 2& 5& 3& 1& 4& 2&6&3&4&5\\
    4& 2& 3& 4& 1& 6& 2& 5& 3&1&5&6&1\\
    5& 6& 1& 5& 2& 4& 3& 1& 6&2&4&3&2\\
    \hline
  \end{array}
  \  \\
n=9 & n=13
\end{array}
$$

$$
\begin{array}{c}
  \begin{array}{|cccccccccccccc|}
    \hline
    1& 3& 2& 6& 3& 1& 5& 2& 4& 3& 2& 5& 6 &4\\
    2& 4& 5& 1& 4& 2& 6& 3& 5& 6& 4& 1& 3& 5\\
    3& 1& 6& 2& 5& 3& 1& 4& 2& 1& 5& 2& 4& 6\\
    4& 2& 3& 4& 1& 6& 2& 5& 3& 4& 6& 3& 5& 1\\
    5& 6& 1& 5& 2& 4& 3& 1& 6& 5& 1& 4& 2& 3\\
    \hline
 \end{array}
 \\
 n=14
\end{array}
$$

$$
\begin{array}{c}
  \begin{array}{|ccccccccccccccccccc|}
    \hline
    1& 3& 2& 6& 3& 1& 5& 2& 4& 3& 1& 4& 2& 1 &6 &2 &5 &3 &4\\
    2& 4& 5& 1& 4& 2& 6& 3& 1& 5& 2& 6& 3& 5 &4 &3 &6 &1& 5\\
    3& 1& 6& 2& 5& 3& 1& 4& 2& 6& 3& 1& 4& 6& 1 &5 &2 &4& 6\\
    4& 2& 3& 4& 1& 6& 2& 5& 3& 1& 4& 2& 5& 3& 2 &6 &3 &5 & 1\\
    5& 6& 1& 5& 2& 4& 3& 1& 6& 2& 5& 3& 6& 4& 5 &1 &4 &6 &2\\
    \hline
  \end{array}
  \\
  n=19
\end{array}
$$

An independent set in $T^2_{5,n}$ has at most one vertex
in any column; thus $\alpha(T_{5,n}^2) \leq n$.
Therefore, $\chi(T_{5,n}^2) \geq 5$ by Observation~\ref{Obs-lowerbound}.
It is easy to check
that $\alpha(T_{5,n}^2) < n$ when $n$ is not a multiple of $5$
(and thus $\chi(T_{5,n}^2)>5$) and that $\alpha(T_{5,7}^2) =5$
(and thus $\chi(T_{5,7}^2)\ge\frac{35}{5}=7$).
\end{proof}

Using combinations of the patterns from Theorem~\ref{Theorem-5n}, we get the following:

\begin{corollary}\label{5k}
If $T_{5k,n}=C_{5k} \Box C_n$ with $k\ge 1$ and $n \geq 5$, then
$$\chi (T_{5k,n}^2) \leq \left \{
\begin{array}{ll}
 5 & \quad \mbox{if $n\equiv 0\pmod 5$,}\\
 7 & \quad \mbox{if $n=7$,}\\
 6 & \quad \mbox{otherwise.}\\
\end{array} \right .$$

\end{corollary}

At this point, we are able to prove our main result.

\mbox{}\par

\noindent{\bf Proof of Theorem~\ref{SW-Theorem}.}
By Corollaries~\ref{3k}, \ref{4k}, and \ref{5k},
we already proved that if one of $m, n$ is a multiple of $3$, $4$, or $5$, then Theorem~\ref{SW-Theorem} holds.
By Lemma~\ref{Lemma-rs} and Corollary~\ref{m12n18}, the remaining cases
are $11 \times 11$, $13 \times 13$,  $13 \times 17$ and $17 \times 17$.
Let $K$ be the $7\times 13$ pattern given in Figure~\ref{Figure-K}.
As in the proof of Theorem~\ref{Theorem-m4n3}, we use combinations of $K$ and $K_3$ (the corresponding pattern,
not the complete graph) to obtain an
$m \times 13$ pattern $X$ for $m \in S(7,3)$.
We then use combinations of $X$ and $X'_4$ to obtain an $m \times n$ pattern for $n \in S(13, 4)$.
In this way, we can obtain proper 7-colorings of
$T_{13,13}^2$, $T_{17,13}^2$, and $T_{17,17}^2$.
We simply transpose the $17 \times 13$ pattern to get a $13 \times 17$ pattern.
Finally, an $11 \times 11$ pattern that properly 6-colors $T^2_{11,11}$ is as follows:
$$
  \begin{array}{|ccccccccccc|}
   \hline
    1& 2& 3& 1& 2& 3& 1& 2& 3& 4& 5\\
 3& 4& 5& 6& 4& 5& 6& 4& 1& 6& 2\\
 5& 1& 2& 3& 1& 2& 3& 5& 2& 3& 4\\
 2& 3& 4& 5& 6& 4& 1& 6& 4& 5& 1\\
 4& 5& 6& 1& 3& 5& 2& 3& 1& 2& 3\\
 1& 2& 3& 4& 2& 6& 4& 5& 6& 4& 5\\
 6& 4& 1& 6& 5& 3& 1& 2& 3& 1& 2\\
 3& 5& 2& 3& 1& 2& 6& 4& 5& 6& 4\\
 1& 6& 4& 5& 6& 4& 5& 1& 2& 3& 5\\
 2& 3& 1& 2& 3& 1& 2& 3& 4& 1& 6\\
 4& 5& 6& 4& 5& 6& 4& 5& 6& 2& 3\\
\hline
 \end{array}
$$ \qed

\begin{figure}
$$\begin{array}{c}
K=
  \begin{array}{|ccccccccccccc|}
    \hline
    1& 2& 3& 4& 5& 6& 1& 2& 3& 4& 5& 6& 7\\
    3& 4& 5& 6& 1& 2& 3& 4& 5& 6& 7& 1& 2\\
    5& 6& 1& 2& 3& 4& 5& 6& 7& 1& 2& 3& 4\\
    1& 2& 3& 4& 5& 6& 7& 1& 2& 3& 4& 5& 6\\
    3& 4& 5& 6& 1& 2& 3& 4& 5& 6& 7& 1& 2\\
    6& 1& 2& 3& 4& 5& 6& 7& 1& 2& 3& 4& 5\\
    4& 5& 6& 1& 2& 3& 4& 5& 6& 7& 1& 2& 3\\
    \hline
  \end{array}
\\
\end{array}
$$
\caption{Pattern for Theorem~\ref{SW-Theorem}\label{Figure-K}}
\end{figure}

As we have seen before, the general upper bound of 7 for
$\chi(T_{m,n}^2)$ given in Theorem~\ref{SW-Theorem} can
be decreased for particular values of $m$ and $n$.
We now provide other cases for which this bound can be decreased to 6.

Using combinations of the $11\times 11$ pattern above, we get:

\begin{corollary}
If $T_{m,n}=C_m \Box C_n$ with $m,n\ge 3$ and $m,n \equiv 0 \pmod {11}$, then
$\chi (T_{m,n}^2)\leq 6.$
\end{corollary}

The same bound can be obtained for $T^2_{6,n}$:

\begin{theorem}
If $T_{6,n}=C_6 \Box C_n$ with $n \geq 6$, then $\chi (T_{6,n}^2)=6.$
\label{Theorem-6n}
\end{theorem}

\begin{figure}
$$
\begin{array}{cc}
\ \ L = 
  \begin{array}{|cccc|}
\hline
1&3&6&4\\
2&4&1&5\\
3&5&2&6\\
4&6&3&1\\
5&1&4&2\\
6&2&5&3\\
\hline
  \end{array}
\ \ &
\ \ M = 
  \begin{array}{|ccc|}
\hline
1&3&5\\
2&4&6\\
3&5&1\\
4&6&2\\
5&1&3\\
6&2&4\\
\hline
  \end{array}
\ \ 
\end{array}
$$
\caption{Patterns for Theorem~\ref{Theorem-6n}\label{Figure-LM}}
\end{figure}

\begin{proof}
Let $L$ and $M$ be the patterns given in Figure~\ref{Figure-LM},
which properly $6$-color $T_{6,4}^2$ and $T_{6,2}^2$, respectively.
By Lemma~\ref{Lemma-rs}, we can get a proper $6$-coloring of $T^2_{6,n}$
by using combinations of patterns $L$ and $M$.
\end{proof}

Using combinations of the patterns from Theorem~\ref{Theorem-6n}, we get the following:

\begin{corollary}
If $T_{6k,n}=C_{6k} \Box C_n$ with $k\ge 1$ and $n \geq 6$, then
$\chi (T_{6k,n}^2)\leq 6.$
\end{corollary}

Finally, using Corollary~\ref{5k}
and the lower bound given by Theorem~\ref{CY-Theorem}, we get the following

\begin{corollary}
If $T_{m,n}=C_m \Box C_n$ with $m,n\geq 3$,
then $\chi(T_{m,n}^2)\geq 5$. Moreover, $\chi(T_{m,n}^2)= 5$ if and only if $m,n \equiv 0 \pmod 5$.
\end{corollary}

\section{Discussion}

\begin{table}
\begin{center}
\begin{tabular}{|c|c|}
  \hline
   values of $m$ and $n$&
   $\chi(T_{m,n}^2)$\\  \hline
$m,n\equiv 0\pmod 5$ & 5 \\ \hline
$m=3$, $n\equiv 0\pmod 2$ & 6 \\ \hline
$m=4$, $n\equiv 0\pmod 3$ & 6 \\ \hline
$m=6$, $n\ge 6$ & 6 \\ \hline
$m=8$, $n=11, 13$ $(*)$ & 6 \\ \hline
$m\equiv 0\pmod 3$, $m\not\equiv 0\pmod 5$, & \\
$n\equiv 0\pmod 2$, $n \not\equiv 0\pmod 5$ & $6$ \\ \hline
$m\equiv 0\pmod 5$, $n\not\equiv 0\pmod 5$, $n\ge 6$, $n\neq 7$ & $6$ \\ \hline
$m\equiv 0\pmod 6$, $n\ge 6$, $n\not\equiv 0\pmod 5$ & $6$ \\ \hline
$m,n\equiv 0\pmod{11}$, $m\not\equiv 0\pmod 5$, $n\not\equiv 0\pmod 5$ & $6$ \\ \hline
$m=3$, $n\not\equiv 0\pmod 2$, $n\neq 3,5$ & 7 \\ \hline
$m=4$, $n\not\equiv 0\pmod 3$, $n\neq 4$ & 7 \\ \hline
$m=5$, $n=7$ & 7 \\ \hline
$m=7$, $n=7,8$ $(*)$ & 7 \\ \hline
$m=3$, $n=5$ & 8 \\ \hline
$m=4$, $n=4$ & 8 \\ \hline
$m=3$, $n=3$ & 9 \\ \hline
\end{tabular}
\end{center}
\caption{\label{Table-Summary}Summary of results on $\chi(T_{m,n}^2)$}
\end{table}

In this paper, we  investigated the chromatic number of the squares of
toroidal grids; that is, squares of Cartesian products of two cycles.
We obtained general upper bounds for this parameter by providing explicit
proper colorings based on the use of specific patterns. This leads in an obvious way
to a linear time algorithm for constructing such colorings.

Table~\ref{Table-Summary} summarizes those of our results that
give tight bounds. 
We also included two cases, marked by $(*)$, for which the tight bound has
been obtained by a computer program.
In all the cases for which a tight bound has
been obtained, this bound matches the lower bound
given by Observation~\ref{Obs-lowerbound}. Therefore, we propose the following:

\begin{conjecture}
For every toroidal grid $T_{m,n}$ with $m,n\ge 3$,
$\chi(T_{m,n}^2) =\Big\lceil \frac{|V(T_{m,n}^2)|}{\alpha(T_{m,n}^2)} \Big\rceil$.
\end{conjecture}

Moreover, it is likely that the chromatic number of the squares of sufficiently large
toroidal grids is at most 6. We therefore propose the following:

\begin{conjecture}
\label{conj2}
The exists some constant $c$ such that for every toroidal grid $T_{m,n}$ with $m,n\ge c$,
$\chi(T_{m,n}^2)\le 6$.
\end{conjecture}

\noindent
{\bf Acknowledgments.}
This work has been done while the second author was visiting the LaBRI thanks
to a postdoctoral fellowship from Bordeaux~1 University.
The first author has been partially supported
by the ANR Project GraTel (Graphs for Telecommunications), ANR-blan-09-blan-0373-01,
2010-2012.\\
We thank the editor for his helpful comments on our manuscript.
Conjecture~\ref{conj2} was suggested to us by the editor and one of the anonymous referees.

\end{document}